\documentclass[11pt,twoside]{article}
\usepackage{asp2014}

\aspSuppressVolSlug
\resetcounters

\begin{document}

\title{UltraPINK - New possibilities to explore Self-Organizing Kohonen Maps}

% full name: Fenja Kollasch
\author{Fenja~Kollasch$^1$ and Kai~Polsterer$^1$}
\affil{$^1$HITS gGmbH, Heidelberg, Germany \email{fenja.kollasch@h-its.org}}
% remove/add as you need

% remove/add authors as you need
\paperauthor{Fenja~Kollasch}{fenja.kollasch@h-its.org}{}{HITS gGmbH}{AIN}{Heidelberg}{Baden-Wuerttemberg}{69118}{Germany}
\paperauthor{Kai~Polsterer}{kai.polsterer@h-its.org}{}{HITS gGmbH}{AIN}{Heidelberg}{Baden-Wuerttemberg}{69118}{Germany}
% remove/add as you need

% leave these next few aindex lines commented for the editors to enable them. Use Aindex.py to generate them for yourself.
% first presenting author should be the first entry for bold-facing the author index page-reference
%\aindex{Kollasch,~F.}
%\aindex{Polsterer,~K.}
% remove/add as you need

% leave the ssindex lines commented for the editors to enable them, use Index.py to suggest yours
%\ssindex{web!application}
%\ssindex{software!infrastructure}
%\ssindex{algorithm!machine learning!unsupervised}

% leave the ooindex lines commented for the editors to enable them, use ascl.py to suggest yours
%\ooindex{FOOBAR, ascl:1101.010}
  
\begin{abstract}
Unsupervised learning algorithms like self-organizing Kohonen maps are a promising approach to gain an overview among massive datasets.
With UltraPINK, researchers can train,
inspect, and explore self-organizing maps, whereby the toolbox of interaction
possibilities grows continually. 

Key feature of UltraPINK is the consideration of versality in astronomical data. By keeping the operations as abstract as possible and using design patterns meant for abstract usage, we ensure that data is compatible with UltraPINK, regardless of its type, formatting, or origin.
Future work on the application will keep extending the catalogue of exploration tools and the interfaces towards other established applications to process
astronomical data. Ultimatively, we aim towards a solid infrastructure for data
analysis in astronomy.
  
\end{abstract}

\section{Introduction}

Data in astronomy is complex on many different levels. As \citet{zhang15} point out,  astronomic data addresses 4 out of the ``10 V'' of big data: 
\begin{enumerate}
	\item The massive grow of data \textit{volume} due to constantly improving observation techniques.
	\item The data \textit{variety} that ranges from image over spectral data towards time series and simulated data.
	\item The \textit{velocity} of data production and analysis that, much like the volume, increases with the technological progress.
	\item The data \textit{values} that come in large scales and usually contain big values.
\end{enumerate}
This circumstance is a medal with two sides:  For once, it helps us to a better comprehenson of astronomic phenomena, but on the other hand, it makes it harder to understand the data itself. This issue has been widely addressed by introducing machine learning concepts to astronomy \citep{kremer17}.

Although this is a promising approach, machine learning is not meant to be an out-of-the-box solution. Automatic data processing helps separating the wheat from the chaff, but in many cases the data still needs manual exploration by researchers. 

This work introduces \textit{UltraPINK}\footnote{Available on GitHub: \url{https://github.com/SirrahErydya/UltraPINK}}, a web application that combines automatic data processing via unsupervised machine learning and manual exploration of the simplified data by relying on generic methods. We will discuss the general concept behind this application and describe its current capabilities.

\section{UltraPINK - An exploration framework for Self-Organizing Kohonen Maps}
With \textit{UltraPINK}, we give an example on how a data exploration pipeline supported by machine learning can be structured. \textit{UltraPINK} is not meant to replace any existing applications, but to introduce the idea of a generic framework. 

As the initial situation, we imagine a data set with many entries such as a survey of galactical images, and a researcher who is looking for none-specific points of interest. \textit{UltraPINK} displays a brief overview among the data by performing dimensionality reduction and allows the researcher to interact with the simplified view in several different ways.

\subsection{Frontend for PINK}
The underlying technology is the \textit{Parallelized rotation and flipping invariant Kohonen maps} framework (\textit{PINK}) \citep{polsterer19}. A self-organizing Kohonen map (\textit{SOM}) is a neural network implementation proposed by \citet{kohonen89}. Through unsupervised learning, a \textit{SOM} expresses the data in a lower dimension. Each neuron of the output layer describes a map cell. After being trained, every cell contains a prototypical representation of a common shape in the data set. Instead of browsing through all the entries, the researcher can examine the \textit{SOM} to get an initial idea about what the data contains.

\textit{UltraPINK} delivers a graphical interface to train, manage, and inspect \textit{SOMs}. As a Django web application \citep{django}, it can be deployed on a local machine for single usage or on a web server to allow multiple people working on the same projects. \textit{PINK} is embedded to the web application via the provided Python package. For each topic the user likes to address, they can create a project in \textit{UltraPINK} and add one or more data sets to this project. On each data set, they can train a \textit{SOM} and proceed to further investigation of the map.

\subsection{Abstract analysis pipeline}
What started as a simple visualization for \textit{SOMs} generated with \textit{PINK} is now the first approach of creating an infrastructure for data exploration combined with machine learning. \textit{UltraPINK} advances its purpose as a frontend by allowing explorative insights into the data, as we proposed in earlier works \citep{O1-98_adassxxx}. Signature of \textit{UltraPINK} is the abstract design pattern on which the interactions are implemented. How an interaction is executed is highly individual and depends on data and research goals. On a higher level however, the definition of interactions is generic: Functions like inspecting, scrolling, zooming, or labelling can be applied to several data sets, regardless of their type. This design decision should prevent us from creating a piece of software that neglects the variety of astronomic data. We are aiming at delivering a high-level pipeline, while still maintaining individual solutions for concrete scenarios.

\section{Features}
Besides mandatory functions that offer the creation and visualization of \textit{SOMs}, \textit{UltraPINK} provides a set of features to enable further investigations and operations on the map or the data itself. The implementation of these operations currently focuses on image data. Due to the mentioned abstract design pattern however, it comes with ease to transfer them to other data types.

\subsection{Inspection and Interaction}
Every trained \textit{SOM} can be inspected in a seperate page providing a toolbox of basic functionalities.  In the following, we describe the most relevant operations.

\paragraph{Selection} 
Each map cell can be selected by mouse click. This results in the enlargment of the respective prototype and the depiction of further information to it. Users can select individual prototypes or multiple at once. Both options are relevant for following actions to be performed.

\paragraph{Show best-matching data points}
When a single prototype is selected, \textit{UltraPINK} gives the option to show the data points resembling this prototype the most. To access this function, a mapping between prototypes and data points has to be calculated. This procedure determines the distance for each prototype to all data points by a given metric. Users can decide on how many data points they want to have displayed. If they select to see the $n$ best fits to a prototype, they receive the $n$ data points with the shortest distance to the prototype.

\paragraph{Show outliers}
Data points with extraordinarily long distances to all prototypes are treated as outliers. Similar to the previous operations, users can have \textit{UltraPINK} display a given number of outliers.

\paragraph{Labeling}
Prototypes, as well as data points can be labeled with \textit{UltraPINK}. Selecting one or more prototypes opens the opportunity of assigning a label. When inspecting the best-matching data points of a prototype, the prototype label can be applied to all data points. It is also possible to give the data points a custom label.

\paragraph{Change of View}
The standard view of the \textit{SOM} inspection site is the depiction of all prototypes corresponding to the map cells. Users can change this view into two other possibilities. The heatmap view deptics the \textit{SOM} as a heatmap, where each cell tells how many data points match best to the prototype assigned with the cell. With the label view, the map cells receive a colored background dependent on the label that was assigned to them.

\subsection{Working with the original data}
Some data belongs to a source but is not relevant to create a \textit{SOM}, such as the position, or the identifier. From the algorithm's perspective, this is metadata that will not be taken into account during learning. Nevertheless, it can be interesting to access this data when a source is inspected. To prevent this information being ignored when examining source data through a \textit{SOM}, \textit{UltraPINK} comes with the option to add a CSV table with metadata to a dataset. Whenever a \textit{SOM} is trained on a dataset with metadata, the dialogue showing the best-matching sources to a prototype gives the option to inspect specific data points. A click on a source image leads to an inspection page where all given metadata is listed. 

Some parts of the metadata can be used for further exploration. For example, if the metadata contains a position, the inspection page uses \textit{ALADIN Lite} \citep{bonnarel2000} to show the source in the sky atlas with a survey of choice. Furthermore, users can generate a list of spatially close sources to the open point to compare features.

The metadata table has no strict format. To support as many different formats as possible, \textit{UltraPINK} uses the data types from \textit{Astropy} \citep{astropy18}.

\section{Summary and Outlook}
We presented the first version of a web application that offers the creation and inspection of self-organizing maps plus the exploration of the underlying observed data with respect to the \textit{SOM}. \textit{UltraPINK} provides basic interaction definitions on an abstract level for arbitrary data and concrete implementations of these methods for image data. Hereby, we consider \textit{UltraPINK} as a prototype of a framework that allows a datatype-independent exploration cycle supported by machine learning.

The toolbox provided by \textit{UltraPINK} grows continuously. Planned for the next public version are image manipulation function in specific, and the support of other data types than images in general. Furthermore, we are working on views for multi-dimensional data.

Collaboration with other research teams is planned to aim towards creating a portfolio of explorative data analysis tools. \textit{UltraPINK} covers one approach of unsupervised machine learning. Instead of forcing it to be a jack of all trades device in future, we rather want it to be one part of a bigger network including multiple tools that are compatible with each other. The next step in this direction would mean to improve \textit{UltraPINKs} compatibility with existing tools and ensure an error-proof workflow.

 \bibliography{X0-016}

\begin{thebibliography}{}
\expandafter\ifx\csname natexlab\endcsname\relax\def\natexlab#1{#1}\fi
\expandafter\ifx\csname url\endcsname\relax
  \def\url#1{\texttt{#1}}\fi
\expandafter\ifx\csname urlprefix\endcsname\relax\def\urlprefix{URL }\fi
\providecommand{\eprint}[2][]{\url{#2}}

\bibitem[{{Astropy Collaboration et al}(2018)}]{astropy18}
{Astropy Collaboration et al} 2018, \aj, 156, 123. \eprint{1801.02634}

\bibitem[{{Bonnarel} et~al.(2000){Bonnarel}, {Fernique}, {Bienaym{\'e}},
  {Egret}, {Genova}, {Louys}, {Ochsenbein}, {Wenger}, \&
  {Bartlett}}]{bonnarel2000}
{Bonnarel}, F., {Fernique}, P., {Bienaym{\'e}}, O., {Egret}, D., {Genova}, F.,
  {Louys}, M., {Ochsenbein}, F., {Wenger}, M., \& {Bartlett}, J.~G. 2000,
  \aaps, 143, 33

\bibitem[{{Django Software Foundation}()}]{django}
{Django Software Foundation}~django: The web framework for perfectionists with
  deadlines.
  \urlprefix\url{https://www.djangoproject.com/https://worldwidetelescope.org/home/}

\bibitem[{Kohonen(1989)}]{kohonen89}
Kohonen, T. 1989, Self-Organization and Associative Memory: 3rd Edition
  (Berlin, Heidelberg: Springer-Verlag)

\bibitem[{{Kollasch} \& {Polsterer}(2021)}]{O1-98_adassxxx}
{Kollasch}, F., \& {Polsterer}, K. 2021, in ADASS XXX, edited by J.-E. {Ruiz},
  \& F.~{Pierfederici} (San Francisco: ASP), vol. TBD of ASP Conf. Ser., 999
  TBD

\bibitem[{{Kremer} et~al.(2017){Kremer}, {Stensbo-Smidt}, {Gieseke},
  {Pedersen}, \& {Igel}}]{kremer17}
{Kremer}, J., {Stensbo-Smidt}, K., {Gieseke}, F., {Pedersen}, K.~S., \& {Igel},
  C. 2017, IEEE Intelligent Systems, 32, 16

\bibitem[{{Polsterer} et~al.(2019){Polsterer}, {Gieseke}, \&
  {Doser}}]{polsterer19}
{Polsterer}, K.~L., {Gieseke}, F., \& {Doser}, B. 2019, {PINK: Parallelized
  rotation and flipping INvariant Kohonen maps}. \eprint{1910.001}

\bibitem[{{Zhang} \& {Zhao}(2015)}]{zhang15}
{Zhang}, Y., \& {Zhao}, Y. 2015, Data Science Journal, 14

\end{thebibliography}

\end{document}